\documentclass[aps,prl,twocolumn,showkeys,floatfix,tightenlines,amsmath,amssymb,nofootinbib,superscriptaddress]{revtex4-1}
\pdfoutput=1

\setcitestyle{super}

\usepackage{ifthen}

\usepackage[T1]{fontenc}
\usepackage{hyperref}
\hypersetup{colorlinks=true,linkcolor=red,citecolor=red}
\usepackage{graphicx}
\usepackage{amssymb}
\usepackage{xcolor}
\usepackage{amsthm}
\usepackage{psfrag}
\usepackage{ifsym}
\usepackage{epstopdf}
\usepackage{dsfont}
\usepackage{amsfonts}
\usepackage{multirow} 
\usepackage{amsmath}
\usepackage{cleveref}

\usepackage{array}
\usepackage{qcircuit}

\newcommand{\ket}[1] {| #1 \rangle}
\newcommand{\braket}[2] {\langle #1 | #2 \rangle}

\newcommand{\one}{\leavevmode\hbox{\small1\normalsize\kern-.33em1}}

\newcommand{\be}{\begin{equation}}
\newcommand{\ee}{\end{equation}}

\begin{document}
\title{Multi-objective evolutionary algorithms for quantum circuit discovery}

\author{V\'aclav Poto\v cek}
\affiliation{Czech Technical University in Prague, Faculty of Nuclear Sciences and Physical Engineering,\\ Prague, Czech Republic}
\affiliation{School of Mathematical and Computer Sciences, Heriot-Watt University, Edinburgh EH14 4AS, UK}

\author{Alan P. Reynolds}
\affiliation{School of Mathematical and Computer Sciences, Heriot-Watt University, Edinburgh EH14 4AS, UK}

\author{Alessandro Fedrizzi}
\affiliation{Scottish Universities Physics Alliance (SUPA), Institute of Photonics and Quantum Sciences, School of Engineering and Physical Sciences, Heriot-Watt University, Edinburgh EH14 4AS, UK}

\author{David W. Corne}
\affiliation{School of Mathematical and Computer Sciences, Heriot-Watt University, Edinburgh EH14 4AS, UK}

\begin{abstract}
Quantum hardware continues to advance, yet finding new quantum algorithms --- quantum software --- remains a challenge, with classically trained computer programmers having little intuition of how computational tasks may be performed in the quantum realm. As such, the idea of developing automated tools for algorithm development is even more appealing for quantum computing than for classical. Here we develop a robust, multi-objective evolutionary search strategy to design quantum circuits `from scratch', by combining and parameterizing a task-generic library of quantum circuit elements. When applied to `ab initio' design of quantum circuits for the input/output mapping requirements of the quantum Fourier transform and Grover's search algorithm, it finds textbook circuit designs, along with alternative structures that achieve the same functionality. Exploiting its multi-objective nature, the discovery algorithm can trade off performance measures such as accuracy, circuit width or depth, gate count, or implementability --- a crucial requirement for first-generation quantum processors and applications.

\end{abstract}


\maketitle

\section{Introduction}
While quantum hardware has progressed rapidly, with access to small-scale quantum processors of upwards of a dozen quantum bits now becoming publicly available, a lack of useful quantum software for these first-generation processors is becoming a barrier to the practical application of quantum computing. Few quantum algorithms with significant speedups over the best classical solution are known to date, and the best known amongst them, such as Shor's factoring algorithm, require thousands of qubits and gate operations to outperform classical approaches. Finding new algorithms has proven to be a very challenging task, due to the lack of a systematic approach to attaining quantum speedup.

In this work we propose using classical, \emph{multi-objective} stochastic optimization methods to help automate quantum circuit discovery. We demonstrate that, knowing only inputs and desired outputs on a fixed number of qubits, a multi-objective genetic algorithm, wrapped around an open-source quantum circuit simulator, not only finds known textbook solutions for Grover's search \cite{grover1996fast} or the quantum Fourier transform \cite{shor1999polynomial} from scratch, but also variants that display trade-offs between accuracy and circuit simplicity.  

In contrast to other approaches to circuit synthesis\cite{dawson2006solovay, kliuchnikov2013fast, bocharov2013efficient,giles2013exact, bocharov2015efficient}, where unitary transformations are expressed as quantum circuits via deterministic decomposition into a fixed set of basic building blocks, stochastic optimization methods such as genetic algorithms work by testing candidate solutions against a prescribed `fitness' criterion and randomly modifying the better solutions in the hope of further improvements. One advantage of this approach is greater freedom in specifying the means of circuit construction, i.e. the permitted quantum gate types and their practical restrictions. Another advantage is complete freedom in the specification of the goal. We need not fix a specific unitary transformation to be performed, if some other way of measuring circuit quality can be found. We may choose to prefer a shorter circuit at the cost of a small increase in error. Moreover, it may not even be necessary to provide a single objective to optimize; a multi-objective approach could be taken using multiple, possibly conflicting, measures of circuit quality.

\begin{figure}[htb]
\centering
\includegraphics[page=1]{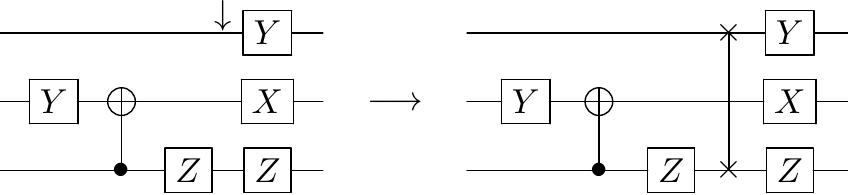}

\vspace{1.2cm}

\includegraphics[page=2]{figs}
\caption{Mutation (gate addition) and crossover applied to quantum circuits. Crossover mimics sexual reproduction, combining `genetic material' from two (or more) parents. Note that the two-point crossover illustrated can be applied to pairs circuits of unequal lengths.}
\label{operators}
\end{figure}

Genetic algorithms are a set of optimization methods inspired by natural selection. A population of (initially random) solutions is evolved through a sequence of \emph{selection}, \emph{reproduction} and \emph{survival} steps. The \emph{selection} step chooses solutions to become `parents'. The \emph{reproduction} step applies randomly chosen operators (see fig.~\ref{operators}) to the parents to create `child' solutions. Finally, the \emph{survival} step chooses which solutions should survive to the next generation. Each solution created is assessed for its fitness according to one or more target objectives and, by promoting the best specimens in the selection and survival steps, the algorithm simulates natural selection, allowing the population to evolve towards better solutions. Genetic algorithms are particularly well suited for complex and rugged parameter landscapes and have been successfully demonstrated for problems in a wide range of research areas, including scheduling \cite{evolutionaryScheduling}, building design \cite{efficiencyAndComfort}, power-distribution networks \cite{generationPlacement} and fault diagnosis \cite{faultDiagnosis}, among many others. In classical electronics, genetic algorithms have been shown to be applicable to automated circuit design \cite{zebulum2001evolutionary,aly2015analog}. It is therefore only natural to apply them to quantum algorithm discovery.

Genetic algorithms have been applied to quantum circuit discovery before, on a limited scale \cite{massey2004evolving, massey2006human, automatedDesign, approxQuantumAdders, autoencoders}, and to the related problem of learning a unitary transformation without decomposing it into a circuit\cite{GAForDeutsch, GAToLearnTransformation}. Also related to our work is research performed on using genetic algorithms to improve digital quantum simulations \cite{GAsAndQuantumSimulation} and on using differential evolution in quantum control \cite{hardQuantumControl, toffoliViaQuantumControl}. Our approach improves upon existing applications of genetic algorithms to circuit discovery, partly thanks to now readily available quantum circuit simulators around which an evolutionary algorithm can be wrapped, but more importantly through the use of a \emph{multi-objective} approach. While search algorithms usually optimize a single objective, many strategies exist for adapting them to multi-objective scenarios \cite{deb, MOTutorial}. These are used whenever the quality of a solution cannot be simply distilled into a single objective function and when it is advantageous to present the user with a range of solutions, representing different trade-offs between multiple objectives. This is the case with quantum circuit discovery: while an indisputable objective is to produce `correct' solutions with negligible theoretical error, it is also important to minimize the number of quantum gates used. There are a number of reasons for this. First, we wish to discover solutions to computational tasks that have lower time complexity (at least asymptotically) than classical approaches. Second, long sequences of quantum operations increase the cumulative error and therefore overheads required for error correction. Also, the physical platform used may imply additional cost criteria, e.g. the number of uses of a particular, hard to implement, gate type. Other possible objectives, not considered by the experiments described here, include the number of auxiliary qubits required by the circuit. A multi-objective algorithm attempts to find a \emph{set} of `Pareto-optimal' solutions for such problems, each of which cannot be improved in one objective without reducing the solution quality according to another. Thus, in a single run, such an algorithm can provide a variety of solutions with differing strengths and weaknesses.

Multi-objective approaches may provide additional benefits to stochastic search algorithms. Their use encourages genetic diversity in population based methods and can help to control bloat --- the tendency, in genetic programming \cite{koza}, for solutions to increase in complexity with little improvement in fitness --- by rewarding simpler solutions. However, multi-objective optimization undeniably places a greater demand on the optimization algorithm and its designers. Discovering, for example, a zero error circuit with the fewest --- let's say 20 --- gates, becomes more challenging when the optimization algorithm is also required to simultaneously discover the most accurate circuits that use no more than 19, or 15, or 10. Increasing the number of objectives further can greatly increase the number of `optimal' circuits to find. Hence, in addition to the usual challenges of finding the best solution representation and the best genetic operators, there is the added task of determining the most suitable set of objectives to use. Following a period of experimentation with all aspects of the algorithm, the code available at \url{https://github.com/vasekp/quantum-ga} represents a robust evolutionary framework that seems effective across a range of disparate tasks.

\section{Results}
We tested the evolutionary framework against two problems with known solutions: finding a decomposition of a unitary discrete Fourier transform, which is at the core of Shor's quantum factoring algorithm \cite{shor1999polynomial}, and finding Grover's algorithm for quantum search \cite{grover1996fast}. Details of the evolutionary strategy and the genetic operators applied to candidate circuits are provided in the Methods section of this paper.

Each problem presented to the underlying algorithm is specified by the number of qubits, a prescription of the measures of merit that form the fitness vector of each circuit (with functions for their evaluation) and a set of permitted gate types. The latter is included to provide the option to use a set of gates that is feasible for the hypothetical laboratory realization of the resulting circuits, not to simplify the algorithm's work.

In the former problem (\textsc{Fourier}), the goal was for the computer to discover, with minimal assumptions, a quantum analogue of the Fast Fourier Transform algorithm\cite{shor1999polynomial}. Thus the algorithm was asked to evolve a circuit that would transform any input quantum superposition to a superposition whose amplitudes are related to those of the input by the discrete Fourier transform, up to a global phase factor. Thanks to the linearity of quantum mechanics it is enough to check this condition for $2^n$ basis states for each candidate circuit, where $n$ is the number of qubits.

The gate set consisted of single-qubit Pauli rotations and arbitrarily controlled phase gates (of arbitrary phase). To simplify comparison of the result with Shor's original algorithm, Pauli $Y$ rotations were chosen. For the same reason the gate set was augmented by a swap gate. Without this addition, the evolution could always find decompositions of the swap operation into rotations and two-qubit gates, but the resulting circuits were difficult to interpret and obtained at the cost of additional running time.

The fitness criteria used were a measure of the overall error with respect to the ideal unitary transform, the worst-case error over the $2^n$ basis states, and the number of instances of each gate type.

\begin{figure}[htb]
\centering
\includegraphics[width=8.6cm]{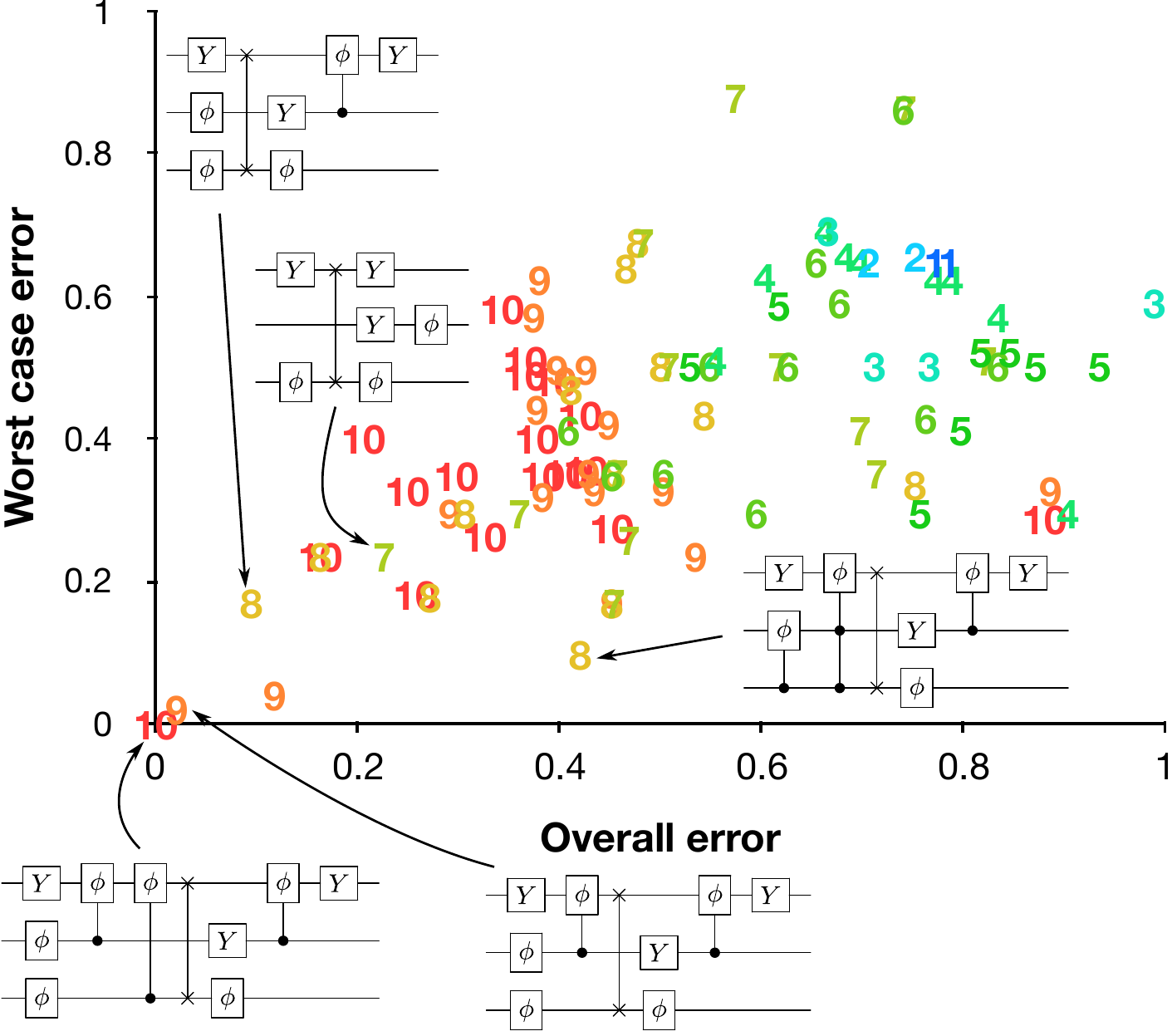}  
\caption{Fitness values of non-dominated solutions found for the \textsc{Fourier} problem in a typical run, with some example circuits. The numbers plotted indicate the number of gates in the circuit, with circuits of more than ten gates omitted. In the example circuits, the angle paremeters for the $Y$ and $\phi$ gates are not plotted.}
\label{FourierResults}
\end{figure}

Initial experiments attempted to solve the 3-qubit problem, for which the standard circuit achieves zero error with 10 gates. 100 runs of the algorithm were performed, of 3000 generations each. (On a standard 4-core $3\,\text{GHz}$ Intel i5-based personal computer, one generation took $21.1 \pm 2.3\,\mathrm{ms}$ to compute in this sample, so 3000 generations correspond to a running time of roughly one minute.) 98 runs found solutions with both worst case and overall errors below $10^{-3}$; on average this was achieved in generation 1053. 92 runs produced circuits that met these error bounds using only 10 gates. On average, runs that met this goal did so in $1358$ generations. Inspection of the non-dominated solutions at termination suggest that the two unsuccessful runs only needed more time --- both had found the 10-gate circuit with errors well below $10^{-2}$. Fig.~\ref{FourierResults} shows the fitness values of solutions in the non-dominated set found in a typical run, and some example circuits.

Note that it is possible to obtain low error values using only 9 gates --- one fewer than that required by the correct circuit. As might be expected, this circuit is the same as the optimal circuit, but with the phase shift gate between qubits 1 and 3 dropped, i.e. the phase gate with the lowest phase change (of $\pi/4$ in this case). Note that merely omitting this gate from the perfect circuit results in an overall error of 0.0565 and a worst case error of 0.0761. Some adjustment of the angle parameters for the remaining gates is required to get these error values down to 0.0193 and 0.0211.

Similar results were obtained for the 4 qubit case. Out of 30 runs of 10000 generations each, a solution with both errors below $10^{-3}$ was obtained in 22, while a solution that met this error bound using the minimal set of 16 gates was found 8 times. Again, low error values (e.g. 0.0049, 0.0051) could be found using 15 gates, by omitting the phase shift gate between qubits 1 and 4 ($\pi/8$) from the perfect solution and adjusting the remaining angle parameters. (Omitting the gate without adjusting the angle parameters produces errors of 0.0144 and 0.0192).

For the \textsc{Grover} problem, we are supplied with an oracle gate that `marks' a single basis state $\ket{x}$ by transforming it to $-\ket{x}$ while leaving all other basis states unchanged. We are not told the identity of $x$. We must embed use of the oracle gate within a quantum circuit so that, initialized by a fixed state, the probability of measuring the output as $x$ (in the standard basis) is maximized, regardless of what $x$ is marked by the oracle. The gate set was similarly restricted to single-qubit (here Pauli $X$) rotations, arbitarily controlled phase gates (of arbitrary phase) and the oracle. To evaluate the circuit, it must be simulated using oracle gates for each of the $2^n$ values of $x$ --- in this sense, the circuit no longer describes a fixed sequence of quantum gates that result in a global unitary operation, but rather a template for constructing one.

Objectives were similar to those used for \textsc{Fourier}: a measure of overall error, the worst-case error, the number of oracle gates used, and the numbers of other gates by type. In the original work \cite{grover1996fast} an application of the oracle gate is equated to an expensive query to a database and the number of oracle gates is taken as a measure of complexity of the algorithm. Here, however, we also endeavour to keep the counts of the other gate types as low as possible, while simultaneously striving for near-zero error.

\begin{figure}[htb]
\centering
\includegraphics[width=8.6cm]{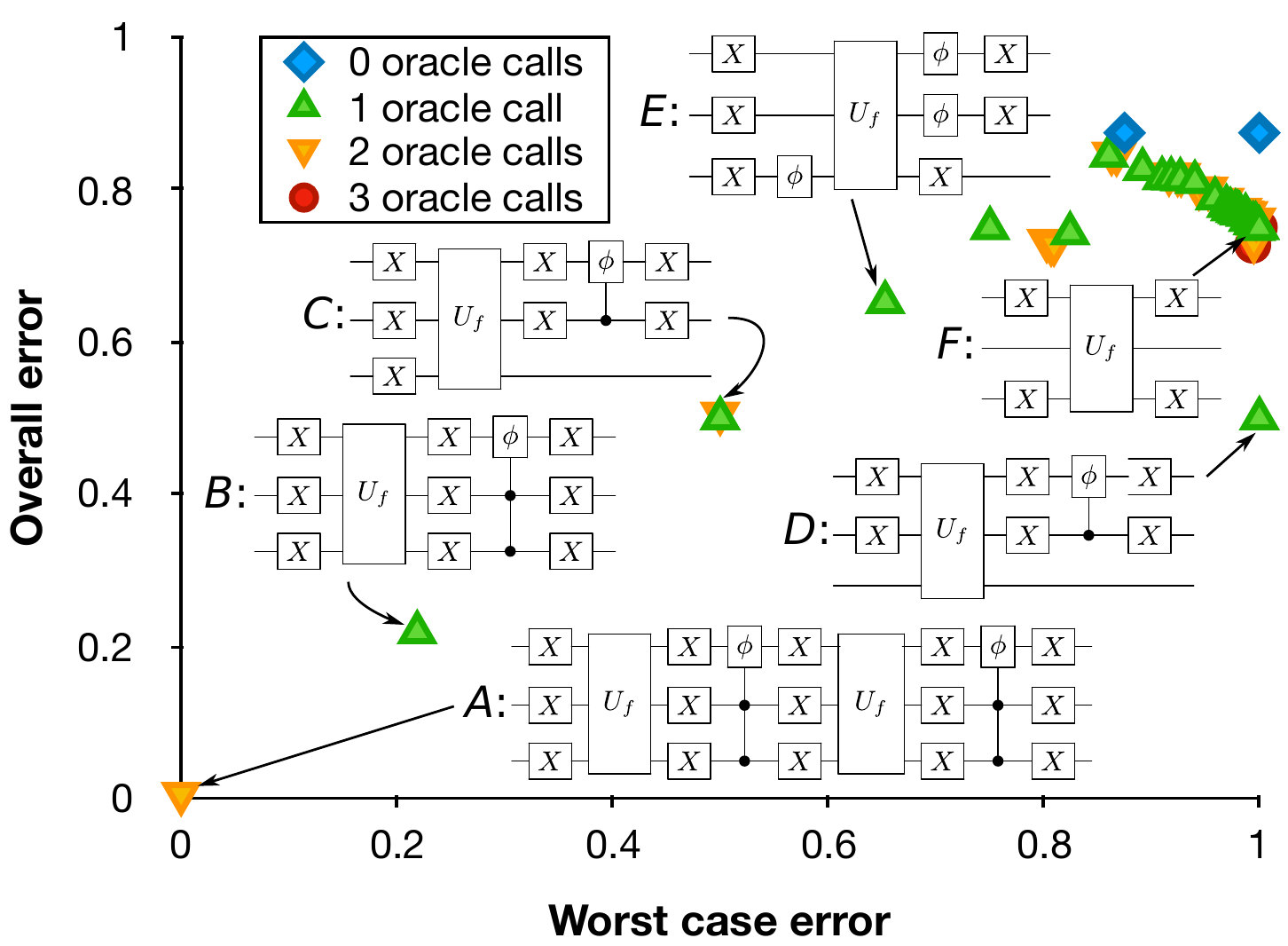}
\caption{Non-dominated solutions, plotted according to accuracy and number of oracle calls, from one of the better runs for the \textsc{Grover} search problem. These include the known optimal solution (circuit $A$) and a truncated version (circuit B).}
\label{groverResults1}
\end{figure}

\begin{figure}[htb]
\centering
\includegraphics[width=8.6cm]{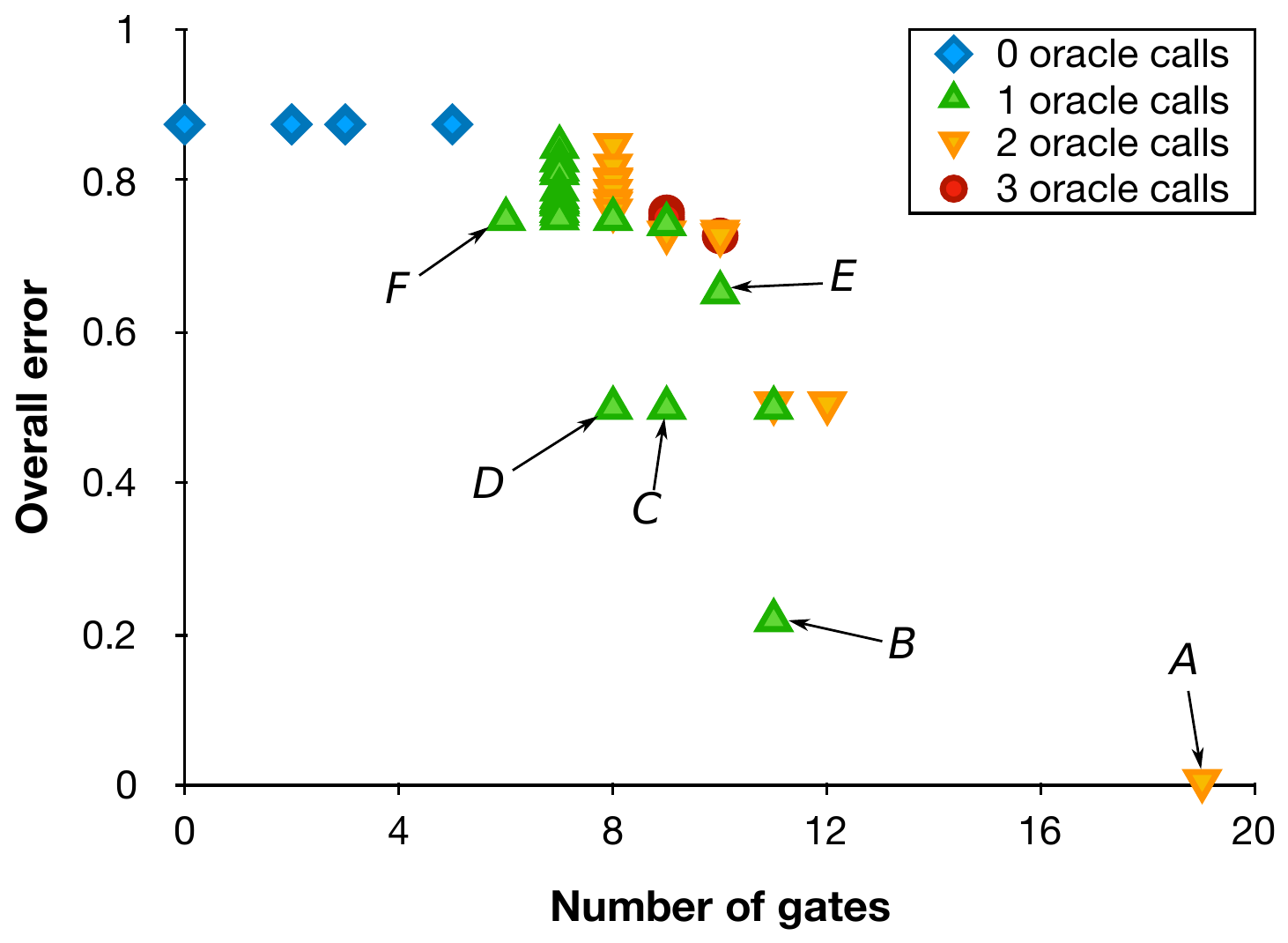}
\caption{The solutions of figure~\ref{groverResults1}, but plotted according to overall accuracy, number of gates and number of oracle calls.}
\label{groverResults2}
\end{figure}

At first, \textsc{Grover} seems more challenging for the discovery algorithm than \textsc{Fourier}: for the 3-qubit problem, only 46 out of 100 runs (using the same parameters as for \textsc{Fourier} --- see Methods) beat $10^{-2}$ in both error measures within 3000 generations (each of $20.8 \pm 5.3\,\mathrm{ms}$), taking, on average, $1036$ generations, to do so. (Compare with \textsc{Fourier} beating the tenfold tighter bound around generation 1053). However, when this target is achieved, it is typically exceeded, with 40 runs giving errors smaller than $10^{-3}$ (typically after 1567 generations). 27 runs (using 1685 generations on average) also met these more stringent error bounds using no more than 2 oracle calls and 19 gates overall --- the size of the known optimal circuit. Non-dominated solutions from a run producing an optimal circuit are shown in figs.~\ref{groverResults1} and~\ref{groverResults2}. 

While the results for \textsc{Grover} might initially seem disappointing, note that Grover's original circuit\cite{grover1996fast} is not perfect either, producing an overall and worst case error of 0.0547, i.e. worse than the targeted error bounds. While modifications to Grover's algorithm have been suggested that produce zero error \cite{GroverZero, amplification, arbitraryPhases}, our approach finds these improved solutions without requiring any extra gates: Grover's original circuit suffices provided the angle parameters of some of the gates, in particular those for the controlled phase shift gates, are adjusted.

The difficulty that the algorithm had in discovering the most accurate circuits may also be partly explained by examining fig.~\ref{groverResults2} --- solution $A$ may only be reached from the other non-dominated solutions through the addition of seven or more gates, making it challenging to find.

Interestingly, our approach appears to find the 4-qubit problem easier to solve. In 30 runs of 10000 generations, errors of less than $10^{-2}$ were obtained in 29 (requiring fewer than $3000$ generations in 27 cases and 1385 generations on average), while errors below $10^{-3}$ were obtained in 26 runs. 10 runs produced circuits with just three oracle calls that satisfied the first error bound, while 5 resulted in three-oracle circuits satisfying the more stringent bound on error. 5 of these 10 runs also used 34 gates --- the number required by the canonical 4-qubit Grover circuit. The discovered three-oracle circuits included some interesting trade-offs, with a number of solutions trading two Pauli $X$ rotations for phase rotations, with no increase in error, and another simply dropping two $X$ rotations from the optimal circuit for errors of 0.0050 and 0.0055.

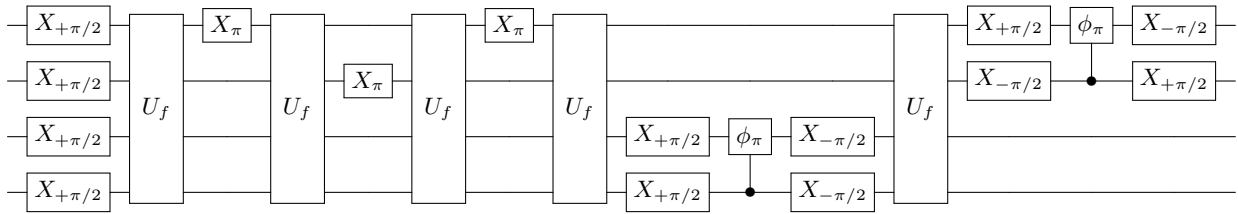
\begin{figure*}
\centering
$
\Qcircuit @C=0.8em @R=.7em
{
& \gate{X_{+\pi/2}} & \multigate{3}{U_f} & \gate{X_{\pi}} & \multigate{3}{U_f} & \qw & \multigate{3}{U_f} & \gate{X_{\pi}} & \multigate{3}{U_f} & \qw & \qw & \qw & \multigate{3}{U_f} & \gate{X_{+\pi/2}} & \gate{\phi_{\pi}} & \gate{X_{-\pi/2}} & \qw\\
& \gate{X_{+\pi/2}} & \ghost{U_f} & \qw & \ghost{U_f} & \gate{X_{\pi}} & \ghost{U_f} & \qw & \ghost{U_f} & \qw & \qw & \qw & \ghost{U_f} & \gate{X_{-\pi/2}} & \ctrl{-1} & \gate{X_{+\pi/2}} & \qw\\
& \gate{X_{+\pi/2}} & \ghost{U_f} & \qw & \ghost{U_f} & \qw & \ghost{U_f} & \qw & \ghost{U_f} & \gate{X_{+\pi/2}} & \gate{\phi_{\pi}} & \gate{X_{-\pi/2}} & \ghost{U_f} & \qw & \qw & \qw & \qw\\
& \gate{X_{+\pi/2}} & \ghost{U_f} & \qw & \ghost{U_f} & \qw & \ghost{U_f} & \qw & \ghost{U_f} & \gate{X_{+\pi/2}} & \ctrl{-1} & \gate{X_{-\pi/2}} & \ghost{U_f} & \qw & \qw & \qw & \qw
}
$
\caption{An interesting 4 qubit circuit for the \textsc{Grover} problem, where the only interactions between the first pair of qubits and the last pair are mediated by the oracle gate. Subscripts give the rotation angles of $X$ and $\phi$ gates. This circuit was found more often than the traditional design, with lower errors.}
\label{grover4Circuit}
\end{figure*}

Excitingly, for four (or more) qubits the evolution also frequently discovers one of a class of novel error-free circuits for Grover's search problem. An example of such a circuit is shown in fig.~\ref{grover4Circuit}.
At the expense of more oracle calls, but with a significant reduction in the overall number of gates, the circuit applies a `divide and conquer' strategy, applying Grover's algorithm to just the bottom two qubits and then to just the top two. Of course, the two-bit Grover's algorithm requires a two-bit oracle, created here in two different ways. First, noting that $X_\pi$ acts (up to overall phase) as a \textsc{Not} gate, the section from the first to the fourth oracle acts as an oracle on just the bottom two bits. 
This two-bit oracle is used to apply Grover's algorithm to the bottom two bits, the output of which is fed into the fifth oracle gate, creating a two-bit oracle for the top two bits. The circuit for the top two bits simplifies to the canonical two-bit Grover's algorithm.

\section{Discussion}
Our results show that multi-objective genetic algorithms are capable of discovering quantum circuits starting with minimal assumptions. The target functionality is not limited to approximating a given unitary transform. This is illustrated by the \textsc{Grover} problem, where an evolved circuit defines a \emph{set} of unitary transformations, one for each type of oracle, while we are only interested in how each of these transformations affects a \emph{single} basis state. The algorithm can use different gates sets while the \textsc{Grover} example illustrates how it can straightforwardly use a non-standard gate type (in this case, the oracle gate) when supplied. Being a \emph{multi-objective} algorithm, it also succeeds in producing a selection of circuits with different trade-offs between objectives such as circuit accuracy, number of gates (either of a particular type, or in total), and potentially other measures of circuit quality such as the number of auxiliary qubits used. Results presented to the user allow for simple exploration of potentially interesting, alternative solutions.

Thus far, we benchmarked our framework against two paradigmatic and dissimilar problems with known solutions. However, even in this setting the genetic algorithm discovered more than we asked for: In the \textsc{Fourier} problem, we found solutions when the \textsc{swap} gate was removed, showing that the genetic algorithm can find equivalent circuits for complex operations with a restricted gate set. For Grover's search, the algorithm discovered novel 4-qubit solutions, and multiple solutions for both problems were found that traded circuit accuracy for simplicity.

Of course, well-established methods exist for the decomposition of target unitary transformations into circuits using a limited gate set\cite{dawson2006solovay, kliuchnikov2013fast, bocharov2013efficient, giles2013exact, bocharov2015efficient}, although in the presence of various experimental restrictions, often only algorithms approaching the known optimal bounds are found. The genetic framework does not aspire to be a direct competitor to these methods, but offers new possibilities, in particular the ability to produce not one but a selection of circuits that trade fidelity for simplicity. An imperfect replacement may deliberately be preferred due to a reduced gate count, reducing the likelihood of an irrecoverable error caused by noise and the need for extremely complicated error correction schemes.

Future improvements to our discovery algorithm will focus on both performance and, more importantly, generalization. With regard to performance, cacheing the results of full or partial solutions has significant potential for improving efficiency, in particular if combined with methods for testing for circuit equivalence. Aborting circuit evaluation as soon as it becomes apparent that the solution is of low quality may reduce the time required per generation, while improved management of the objectives may reduce the number of generations required. The algorithm could also be hybridized with other methods, e.g. numerical optimization of gate parameters for a fixed circuit design.

More important, however, is the potential for generalizing the approach. In the present state of the evolutionary framework, the number of qubits for the quantum circuit is fixed as a part of specification of the problem. In contrast, descriptions of Grover's and Shor's algorithms provide a \emph{recipe} for building circuits, for \emph{arbitrary} sized input. We cannot expect our algorithm to produce Grover's \emph{algorithm}; the best we can hope for from a single run is a \emph{circuit} that implements Grover's algorithm for some fixed input size.  To extend our current approach to find \emph{scalable} quantum algorithms, one would need first to perform optimization runs for different input sizes, then detect a rule connecting the circuits found, and finally provide a proof of the validity of extrapolating this rule to an arbitrary number of qubits.

An alternative to attempting to generalize from discovered small-scale circuits is to search, instead, in the space of rules, or algorithms, for constructing circuits. While it is possible that a genetic algorithm could successfully optimize a recipe for circuit construction, determining how to represent such a recipe to the algorithm and how genetic operators should modify the representation is clearly more challenging, though some initial steps \cite{GPForCircuits, GPForCircuitsBook} in this direction have been taken using genetic programming \cite{koza}.

Finally, given a problem to solve, it is typically not obvious what the input and output of the successful quantum circuit will be. For example, for factorization, it is not immediately clear that a quantum circuit that performs Fourier transforms is required. Quantum algorithm design is about more than matching the outputs of a circuit with some desired output --- typically, a certain amount of mathematical manipulation is required to show (and prove) how the output results in a solution to the problem. Integrating these manual steps into automatic discovery is a formidable but worthwhile challenge.

\section{Methods}
The optimization strategy used is inspired by the NSGA multi-objective evolutionary algorithm \cite{srinivas1994multiobjective}, yet incorporates elitist aspects of SPEA2 \cite{zitzler2002spea2}. The decision to incorporate this elitism, along with others in the algorithm design, is influenced by the need of any effective genetic algorithm to balance exploration, by maintaining the diversity of the gene pool, against exploitation of good solutions via a more focused search. Maintaining a diverse population means that elements of different solutions can be combined via crossover, while keeping the search focused means that these parent solutions are likely to be of high quality.

\subsection{Encoding}
The nature of the problem requires variable-length encoding, in order to effectively describe circuits of different size. While a solution is represented simply as a string of gates, different types of gate require the storage and manipulation of different information: a controlled-NOT gate requires the specification of the control and target qubit indices, while a single bit rotation gate is described by a single qubit index and a rotation angle. Hence the genome is encoded using a shallow data structure, consisting of a sequence of genes, one for each gate, each of which contains information specific to the gate type. Each gene is therefore a pointer to a polymorphic object, encoding both gate type and additional information.

As a result, the genetic algorithm code, when applying an operator to a single gate, sends a message to a gate object indicating the \emph{intent} of the modification, e.g. gate inversion, which is then implemented by the gate object in an appropriate way. For example, the `alterContinuous' message is translated by a rotation gate as a request to adjust the angle parameter, while the swap gate interprets this message as a request to change the target bits.

\subsection{Genetic operators}
The following genetic operators are used in the experiments of this paper. Some are standard operators for variable length genotypes, such as gate insertions, moving a gate within the sequence or mutating a single gate, while others have designed specifically for quantum circuit search. For example, the `sequence and inverse insertion' effects a change of basis over a section of circuit. Using such domain-specific genetic operators has been widely established as advantageous. In addition, a balance has been struck between operators that make minor changes, to facilitate the exploitation of good solutions through a sequence of minor improvements, and more disruptive operators that facilitate exploration of the search space.

The genetic operator used to create a new solution is chosen uniformly at random from the following list.
\begin{description}
\item[1. Discrete, uniform mutation] Mutate selected gates, keeping the existing gate types, but changing target and/or control bits.
\item[2. Continuous, uniform mutation] Mutate selected gates, adjusting a continuous parameter such as rotation angle, where possible. (If not possible, e.g. for swap gates, a discrete mutation is performed instead.)
\item[3. Sequence insertion] Insert a contiguous sequence of randomly generated gates into the genome.
\item[4. Sequence and inverse insertion] Insert a contiguous sequence of randomly generated gates into the genome at one point, and the inverse of this sequence at a later point.
\item[5. Insert, mutate, invert] Perform a discrete mutation on a single gate in the genome, then place a randomly selected gate immediately before it within the genome and the inverse of that gate immediately after.
\item[6. Swap qubits] For a contiguous stretch of the genome, swap the roles of two randomly selected qubits.
\item[7. Sequence deletion] Delete a contiguous sequence from the genome.
\item[8. Sequence replacement] Replace a contiguous sequence of gates with a randomly generated sequence. The new sequence need not be the same length as the original.
\item[9. Sequence swap] Select two (non-overlapping) contiguous sequences of gates within the genome and swap.
\item[10. Sequence scramble] Randomly permute the gates from a contiguous sequence in the genome.
\item[11. Move gate] Move a single gate to a different point in the genome.
\item[12. Crossover] Select two parent solutions and create an empty child. Starting at the beginning of each parent solution, randomly pick the number of gates to be selected. Add the gates selected from the first parent to the child, while discarding those from the second. Repeat, alternating which parent donates gates to the child. (Multi-point crossover.)
\end{description} 

Operations 1, 2 and 12 take a common parameter called the expected mutation count (EMC), while operations 3, 4, 6, 7, 8 and 10 take a similar parameter called the expected sequence length (ESL). In each of our experiments, these parameters are both set to 2.0. For operations 1 and 2, the EMC is simply divided by the genome length, $\ell$, to give the probability of mutation of any gate. Operations 3 and 4 select the sequence insertion point (or points) uniformly at random, while the length of the sequence added is drawn from the geometric distribution with mean given by ESL. Operations 6, 7, 8 and 10 select the start point of the stretch of genome to be modified uniformly at random, then draw the sequence length from the geometric distribution with mean given by ESL, but in these cases there is the potential for reaching the end of the genome, at which point the operation is halted. Operation 9 selects four points in the genome uniformly at random, ensuring that the two sequences so defined include at least one gate each. Finally operation 12 uses a geometric distribution with mean $\ell/\text{EMC}$ to determine the length of each chunk of solution to be copied to the child (or discarded). This results in an expected number of crossover points approximately equal to EMC.

Mutation of gate rotation parameters is performed by adding a value drawn from the normal distribution with zero mean and a standard deviation of 0.2. In the mutation of control and target qubits, each qubit is given a 50\% chance of being a control, then the target is selected uniformly at random.

Other genetic operators have been considered and implemented, but did not provide any appreciable advantage.

\subsection{Dominance and ranking}
As usual in multi-objective algorithms, pairs of circuits are compared using the notion of dominance: circuit $A$ dominates circuit $B$ if the fitness of $A$ is better or equal than that of $B$ in each objective, and strictly better in at least one. This notion of dominance is then used to assign a rank to each candidate in the population, using the non-dominated sorting method of NSGA \cite{srinivas1994multiobjective}. Circuits that are not dominated by any other in the population are assigned a rank of zero. Those that are dominated only by circuits of rank zero are assigned a rank of one, and so on.

We use the adjective `non-dominated' to describe those solutions that are not dominated by any others in the set of solutions under consideration, for example the population. We use `Pareto-optimal' to describe solutions that are not dominated by \emph{any} other solutions.

\subsection{Selection and replacement}
The evolution starts with an initial population of 1000 candidates, with gate types and gate parameters all selected uniformly at random. Control and target bits are selected in the same way as when these properties are mutated, as described above. The length of each genome is selected from the geometric distribution, with the mean currently set to 30 gates.

After the evaluation of the fitness vector for each solution, the population is processed in a number of selection and replacement steps, combining features of the NSGA \cite{srinivas1994multiobjective} and SPEA2 \cite{zitzler2002spea2} algorithms. After ranking solutions according via nondominated sorting, each circuit is assigned a selection probability proportional to $e^{-ar}$, where $r$ is the rank. The parameter $a$ controls the selection pressure and is set to 1 for the experiments. These selection probabilities remain fixed for the duration of a single generation and give the probability that a solution will be chosen whenever a parent solution is required for mutation or crossover.

Control of the survival of solutions in the NSGA algorithm is straightforward: child solutions are created until they equal the original population in number, after which the old solutions are discarded. However, given the disruptive nature of some of the genetic operators, the concern is that high quality solutions --- the result, perhaps, of fortunate mutation --- will be lost to the algorithm soon after their creation. Hence we incorporate some of the elitism of SPEA2 \cite{zitzler2002spea2}. Prior to the creation of child solutions, up to 100 of the best solutions are copied directly into the new generation. The set of solutions to be preserved in this way is obtained by first selecting the non-dominated solutions from the old population. If pairs of solutions are found to differ in fitness values by less than 0.1, using the Manhattan metric to measure distance, then one of the pair is removed. Finally, if there are more than 100 solutions remaining, the best 100 according to the most important objective --- the overall error in our experiments --- are kept. (Note that this is the only part of the algorithm that does not treat the objectives equally.) The remaining 900 (or more) spaces in the new population are then filled with child solutions obtained by applying mutation or crossover operators to parents selected from the old population.

Once the population has been filled, some additional pruning takes place. If two circuits have identical fitness vectors, one is removed. Furthermore, if two circuits are structurally the same, differing only in angle parameters, one is removed. (If, in this situation, one of the circuits dominates the other, then the better solution is kept.) This completes a generation.

Some limited circuit simplification is performed within the algorithm. If two consecutive gates in the genome are of the same type, then they are merged when possible. Note, though, that much more thorough simplification is, of course, possible.

\subsection{Error functions}
There are some subtleties in the calculation of the error functions, which differs somewhat for the two test problems. In both cases, the calculation involves the overlap between the output of the circuit, $\ket{\psi}$ and the desired output, $\ket{\chi}$, i.e. $\braket{\psi}{\chi}$.

In what follows, we assume that we are working with four qubits and we label the basis states from $\ket{0} = \ket{0000}$ to $\ket{15} = \ket{1111}$. The \textsc{Grover} case is the most straightforward. We are told that one of the basis states is special and we are provided access to an oracle gate that `marks' this special state. Input into the circuit is always $\ket{0} = \ket{0000}$, but a simulation must be performed using each of the 16 possible versions of the oracle gate. Upon measurement of the output (in the standard basis), there should be a high probability that the measurement correctly identifies which of the basis states is special. The error function for one of these 16 runs is simply the complement of the probability that the special state is correctly identified, i.e.
\[
E = 1 - |\braket{\psi}{\chi}|^2.
\] 
The worst case error is then simply the largest of these errors, while the overall error is given by the average of these 16 error values.

For the \textsc{Fourier} problem, the aim is to produce a circuit that performs a quantum Fourier transform on any input. Of course, one need only perform simulations for each of the basis states, requiring 16 runs. The complication is that the circuit should also be considered successful if the output differs from the desired output only by an overall  (unphysical) phase. Hence, for a single input, the error is considered to be
\[
E = 1 - |\braket{\psi}{\chi}|.
\]
The worse case error is then given by the largest of these values. However, when considering the overall error, one must consider that, to be a correct circuit, the outputs must not only match the desired output up to a phase, but this phase must also be the same for each of the 16 inputs. As a result, a suitable expression for the overall error is given by
\[
E_{\text{overall}} = 1 - \sum_{i = 0}^{15}|\braket{\psi_i}{\chi_i}| / 16.
\]
Unlike in the \textsc{Grover} problem, the overall error for the \textsc{Fourier} problem is not an average --- indeed it may be greater than the worst case error.

\subsection{Simulation}
The optimization code has been designed to be integrated with different quantum circuit simulators. Thus far, it works with Quantum++ \cite{quantum++} and QIClib \cite{QIClib}. Experiments were performed using QIClib.

\section{Data Availability}
Source code (tested on linux, using gcc) is available from \url{https://github.com/vasekp/quantum-ga}, or from the corresponding author on reasonable request.

\section{Acknowledgements}
The work reported here was supported by the Networked Quantum Information Technologies (NQIT) project, a `Quantum Hub' led by the University of Oxford, and funded by the EPSRC's UK National Quantum Technologies Programme (grant number EP/M013243/1). This work formed part of an `Industry Partnership' project, with additional support and engagement from Route Monkey Ltd. A.F. also acknowledges support from EPSRC grant number EP/N002962/1.

\section{Competing Interests}
The authors declare no competing interests.

\section{Contributions}
A.F. and D.W.C. initiated this research project. V.P. and D.W.C. co-designed the algorithm. V.P. wrote the genetic algorithm code. V.P. and A.P.R performed and evaluated circuit discovery experiments. V.P. and A.P.R. prepared the manuscript with feedback and assistance from A.F. and D.W.C.

\newpage
\appendix
\onecolumngrid
\clearpage
\renewcommand{\theequation}{S\arabic{equation}}
\renewcommand{\thefigure}{S\arabic{figure}}
\renewcommand{\thetable}{\Roman{table}}
\renewcommand{\thesection}{S\Roman{section}}
\setcounter{equation}{0}
\setcounter{figure}{0}


\begin{thebibliography}{10}
\providecommand{\url}[1]{\texttt{#1}}
\providecommand{\urlprefix}{URL }
\providecommand{\eprint}[2][]{\url{#2}}

\bibitem{grover1996fast}
Grover, L.~K.
\newblock {A fast quantum mechanical algorithm for database search}.
\newblock In \emph{Proceedings of the Twenty-Eighth Annual ACM Symposium on
  Theory of Computing}, STOC '96, 212--219 (ACM, 1996).

\bibitem{shor1999polynomial}
Shor, P.~W.
\newblock {Polynomial-Time Algorithms for Prime Factorization and Discrete
  Logarithms on a Quantum Computer}.
\newblock \emph{SIAM review} \textbf{41}(2), 303--332
\newblock (1999).

\bibitem{dawson2006solovay}
Dawson, C.~M. \& Nielsen, M.~A.
\newblock {The {Solovay-Kitaev} algorithm}.
\newblock \emph{Quantum Information and Computation} \textbf{6}(1), 81--95
\newblock (2006).

\bibitem{kliuchnikov2013fast}
Kliuchnikov, V., Maslov, D. \& Mosca, M.
\newblock Fast and efficient exact synthesis of single-qubit unitaries
  generated by {Clifford} and {$T$} gates.
\newblock \emph{Quantum Information \& Computation} \textbf{13}(7-8), 607--630
\newblock (2013).

\bibitem{bocharov2013efficient}
Bocharov, A., Gurevich, Y. \& Svore, K.~M.
\newblock {Efficient decomposition of single-qubit gates into {$V$} basis
  circuits}.
\newblock \emph{Physical Review A} \textbf{88}(1), 012313
\newblock (2013).

\bibitem{giles2013exact}
Giles, B. \& Selinger, P.
\newblock {Exact synthesis of multiqubit {Clifford+$T$} circuits}.
\newblock \emph{Physical Review A} \textbf{87}(3), 032332
\newblock (2013).

\bibitem{bocharov2015efficient}
Bocharov, A., Roetteler, M. \& Svore, K.~M.
\newblock Efficient Synthesis of Universal Repeat-Until-Success Quantum Circuits.
\newblock \emph{Physical Review Letters} \textbf{114}(8), 080502
\newblock (2015).

\bibitem{evolutionaryScheduling}
Hart, E., Ross, P. \& Corne, D.
\newblock Evolutionary Scheduling: A Review.
\newblock \emph{Genetic Programming and Evolvable Machines} \textbf{6}(2), 191--220
\newblock (2005).

\bibitem{efficiencyAndComfort}
Yu, W., Li, B., Jia, H., Zhang, M. \& Wang, D.
\newblock {Application of multi-objective genetic algorithm to optimize energy
  efficiency and thermal comfort in building design}.
\newblock \emph{Energy and Buildings} \textbf{88}, 135--143
\newblock (2015).

\bibitem{generationPlacement}
Evangelopoulos, V. \& Georgilakis, P.~S.
\newblock {Optimal distributed generation placement under uncertainties based
  on point estimate method embedded genetic algorithm}.
\newblock \emph{IET Generation, Transmission \& Distribution} \textbf{8}(3),
  389--400
\newblock (2014).

\bibitem{faultDiagnosis}
Cerrada, M.
  \emph{et~al.}
\newblock {Fault diagnosis in spur gears based on genetic algorithm and random
  forest}.
\newblock \emph{Mechanical Systems and Signal Processing} \textbf{70--71},
  87--103
\newblock (2016).

\bibitem{zebulum2001evolutionary}
Zebulum, R.~S., Pacheco, M.~A. \& Vellasco, M. M.~B.
\newblock \emph{{Evolutionary Electronics: Automatic Design of Electronic
  Circuits and Systems by Genetic Algorithms}}, vol.~22 of \emph{The CRC Press
  International Series on Computational Intelligence} (CRC press, 2001).

\bibitem{aly2015analog}
Aly, W.~M.
\newblock {Analog Electric Circuits Synthesis using a Genetic Algorithm
  Approach}.
\newblock \emph{International Journal of Computer Applications}
  \textbf{121}(4)
\newblock (2015).

\bibitem{massey2004evolving}
Massey, P., Clark, J. \& Stepney, S.
\newblock Evolving quantum circuits and programs through genetic programming.
\newblock In \emph{Genetic and Evolutionary Computation--GECCO 2004}, 569--580
  (Springer, 2004).

\bibitem{massey2006human}
Massey, P., Clark, J.~A. \& Stepney, S.
\newblock {Human-Competitive Evolution of Quantum Computing Artefacts by
  Genetic Programming}.
\newblock \emph{Evolutionary Computation} \textbf{14}(1), 21--40
\newblock (2006).

\bibitem{automatedDesign}
Williams, C.~P. \& Gary, A.~G.
\newblock {Automated Design of Quantum Circuits}.
\newblock In \emph{Quantum Computing and Quantum Communications}, vol. 1509 of
  \emph{Lecture Notes in Computer Science}, 113--125 (Springer, 1999).

\bibitem{approxQuantumAdders}
Li, R., Unai, A.-R., Lamata, L. \& Solano, E.
\newblock {Approximate Quantum Adders with Genetic Algorithms: An {IBM} Quantum
  Experience}.
\newblock \emph{Quantum Measurements and Quantum Metrology} \textbf{4}(1),
  1--7
\newblock (2017).

\bibitem{autoencoders}
Lamata, L., Alvarez-Rodriguez, U., Martin-Guerrero, J.~D., Sanz, M. \& Solano,
  E.
\newblock {Quantum autoencoders via quantum adders with genetic algorithms}.
\newblock \emph{Quantum Science and Technology} \textbf{4}(1), 014007
\newblock (2019).

\bibitem{GAForDeutsch}
Bang, J. \& You, S.
\newblock {A Genetic-algorithm-based Method to Find Unitary Transformations for
  Any Desired Quantum Computation and Application to a One-bit Oracle Decision
  Problem}.
\newblock \emph{Journal of the Korean Physical Society} \textbf{65}(12),
  2001--2008
\newblock (2014).

\bibitem{GAToLearnTransformation}
Spagnolo, N.
  \emph{et~al.}
\newblock {Learning an unknown transformation via a genetic approach}.
\newblock \emph{Scientific Reports} \textbf{7}(1), 14316
\newblock (2017).

\bibitem{GAsAndQuantumSimulation}
Las~Heras, U., Alvarez-Rodriguez, U., Solano, E. \& Sanz, M.
\newblock Genetic Algorithms for Digital Quantum Simulations.
\newblock \emph{Physical Review Letters} \textbf{116}(23), 230504
\newblock (2016).

\bibitem{hardQuantumControl}
Zahedinejad, E., Schirmer, S. \& Sanders, B.~C.
\newblock {Evolutionary algorithms for hard quantum control}.
\newblock \emph{Physical Review A} \textbf{90}(03), 032310
\newblock (2014).

\bibitem{toffoliViaQuantumControl}
Zahedinejad, E., Ghosh, J. \& Sanders, B.~C.
\newblock {High-Fidelity Single-Shot Toffoli Gate via Quantum Control}.
\newblock \emph{Physical Review Letters} \textbf{114}(20), 200502
\newblock (2015).

\bibitem{deb}
Deb, K.
\newblock \emph{{Multi-Objective Optimization using Evolutionary Algorithms}}
  (John Wiley \& Sons, Ltd., 2001).

\bibitem{MOTutorial}
Konak, A., Coit, D.~W. \& Smith, A.~E.
\newblock {Multi-objective optimization using genetic algorithms: {A}
  tutorial}.
\newblock \emph{Reliability Engineering and System Safety} \textbf{91},
  992--1007
\newblock (2006).

\bibitem{koza}
Koza, J.~R.
\newblock \emph{{Genetic Programming: On the Programming of Computers by Means
  of Natural Selection}} (MIT Press, 2015).

\bibitem{GroverZero}
Long, G.~L.
\newblock {Grover algorithm with zero theoretical failure rate}.
\newblock \emph{Physical Review A} \textbf{64}(2), 022307
\newblock (2001).

\bibitem{amplification}
Brassard, G., H{\o}yer, P., Mosca, M. \& Tapp, A.
\newblock {Quantum amplitude amplification and estimation}.
\newblock In \emph{Quantum Computation and Quantum Information}, vol. 305 of
  \emph{Contemporary Mathematics} (eds. Lomonaco, Jr., S.~J. \& Brandt, H.~E.),
  53--74 (American Mathematical Society, 2002).

\bibitem{arbitraryPhases}
H{\o}yer, P.
\newblock {Arbitrary phases in quantum amplitude amplification}.
\newblock \emph{Physical Review A} \textbf{62}(5), 052304
\newblock (2000).

\bibitem{GPForCircuits}
Spector, L., Barnum, H., Bernstein, H.~J. \& Swamy, N.
\newblock Quantum Computing Applications of Genetic Programming.
\newblock In \emph{Advances in Genetic Programming}, vol.~3 (eds. Spector, L.,
  Langdon, W.~B., O'Reilly, U.-M. \& Angeline, P.~L.), 135--160 (MIT Press,
  1999).

\bibitem{GPForCircuitsBook}
Spector, L.
\newblock \emph{{Automatic Quantum Computer Programming: A Genetic Programming
  Approach}}.
\newblock Genetic Programming Series (Kluwer Academic Publishers, 2004).

\bibitem{srinivas1994multiobjective}
Srinivas, N. \& Deb, K.
\newblock {Multiobjective Optimization Using Nondominated Sorting in Genetic
  Algorithms}.
\newblock \emph{Evolutionary Computation} \textbf{2}(3), 221--248
\newblock (1994).

\bibitem{zitzler2002spea2}
Zitzler, E., Laumanns, M. \& Thiele, L.
\newblock {{SPEA2}: Improving the Strength Pareto Evolutionary Algorithm for
  Multiobjective Optimization}.
\newblock In \emph{Evolutionary Methods for Design, Optimisation and Control
  with Application to Industrial Problems (EUROGEN 2001)} (eds. Giannakoglou,
  K. \emph{et~al.}), 95--100 (International Center for Numerical Methods in
  Engineering (CIMNE), 2002).

\bibitem{quantum++}
Gheorghiu, V.
\newblock {Quantum++: A {C++11} quantum computing library}.
\newblock Preprint at \url{https://arxiv.org/abs/1412.4704}
\newblock (2014).

\bibitem{QIClib}
Chanda, T.
\newblock {The Quantum Information and Computation Library ({QIClib})}
\newblock (2015--2017), \url{https://titaschanda.github.io/QIClib/}.

\end{thebibliography}
\end{document}